\documentclass[12pt,a4paper]{article}

\renewcommand{\title}[1]{
\begin{center} \Large \bf #1 \end{center}
}

\renewcommand{\author}[3]{
 \begin{center} #1 \\
  {\it #2} \\
  {\small E-mail: \texttt{#3}}
 \end{center}
\addvspace{\baselineskip}
}

\usepackage{amssymb}
\usepackage{amsmath}

\usepackage{theorem}
\theoremstyle{break}

\theoremstyle{break}

\theoremstyle{break}

\def\tr{{\hbox{\rm Tr}}}

\begin{document}

\begin{titlepage}

\baselineskip 5mm

\begin{flushleft}
December 2003
\end{flushleft}

\begin{flushright}
KSTS/RR-03/007 \\
hep-th/0312120
\end{flushright}

\title{Noncommutative Cohomological Field Theories and\\
Topological Aspects of Matrix models}

\author
{Akifumi Sako${}^{\dagger}$  \\ \ }
{${}^{\dagger}$ Department of Mathematics, Faculty of Science and
 Technology, Keio University\\
3-14-1 Hiyoshi, Kohoku-ku, Yokohama 223-8522, Japan\\ \  }
{${}^{\dagger}$ sako@math.keio.ac.jp}

\vspace{1cm}

\abstract{
We study topological aspects of matrix models and noncommutative 
cohomological field theories (N.C.CohFT).
N.C.CohFT have symmetry under an arbitrary
infinitesimal deformation with noncommutative parameter $\theta$.
This fact implies that N.C.CohFT are topologically less sensitive 
than K-theory, but the 
classification of manifolds by N.C.CohFT opens the possibility to 
get a new view point for 
global characterization of noncommutative manifolds.
To investigate the properties of N.C.CohFT, we construct some
models whose fixed point loci are given by sets of 
projection operators. In particular, the partition function on the
Moyal plane is calculated by using a matrix model.
The moduli space of the matrix model is a union of Grassman manifolds.
The partition function of the matrix model is calculated 
using the Euler number of the Grassman manifold.
Identifying the N.C.CohFT with the matrix model,
we obtain the partition function of the N.C.CohFT.
To check the independence of the noncommutative
parameters, we also study the moduli space in the 
large $\theta$ limit and for finite
$\theta$, for the case of the Moyal plane.
If the partition function of N.C.CohFT is topological in the
sense of noncommutative geometry,
then this should reveal some relation with K-theory.
Therefore we investigate certain models of CohFT and N.C.CohFT 
from the point of view
of K-theory.
Our observations give us an analogy between CohFT and N.C.CohFT
in connection with K-theory. 
Furthermore, we verify for the Moyal plane and 
noncommutative torus cases that our partition functions are 
invariant under those deformations which do not change the K-theory.
Finally, we discuss the noncommutative cohomological Yang-Mills theory.
}

\end{titlepage}

\section{ Introduction}

Recent developments in string theory provide a fruitful framework 
and motivation for physicists to study noncommutative field theories.
{}From the viewpoint of physics, much progress has been achieved 
by using noncommutative geometry. 
On the other hand, from the point of view of noncommutative space geometry
and topology investigated by physical techniques, 
there are some successful cases, like 
for example Kontsevich's deformation quantization.
It is known that we can construct Kontsevich's deformation quantization
by using 
some kind of topological string theory \cite{Kontsevich,Cattaneo}.
As another example are the investigations of topological charges. The role 
of these charges in the noncommutative case is not completely clear yet, 
however they are topological in commutative space, and thus 
these charges have some kind of topological nature
\cite{sako2,sako3,Furuuchi1,Furuuchi2,Furuuchi3,Furuuchi4,Tian,Kim3,Bak,Dileep}.

Topology and geometry of ``commutative" space are studied by
many methods.  One important way to investigate them is
to use quantum (or classical) field theories and string theories.
For example, Donaldson theory, Seiberg-Witten theory, Gromov-Witten
theory are constructed 
by cohomological field theories(CohFT).
Therefore, it is natural to ask:  
`` Can Noncommutative Cohomological field theories (N.C.CohFT) 
be used for the investigation of noncommutative geometry or topology?" 
Here we call N.C.CohFT the CohFT which is naively extended to noncommutative 
space. 
One of the aims of this article is to give the circumstantial evidence 
for a positive answer to this question.

Noncommutative spaces are often defined by using an algebraic formulation, 
for example by using $C^* $ algebras. 
So their topological discussions are usually based on algebraic K-theory.
For example, the rank of $K_0$ 
identifies each noncommutative torus $T^2_{\theta}$ that is characterized
by the noncommutative parameter $\theta$.
In this sense, even if $\theta - \theta'$ is arbitrary small,
$T^2_{\theta}$ is distinguished from $T^2_{\theta'}$ without 
Morita equivalent cases.
Meanwhile, some topological charges 
in commutative space seem to remain ``topological"
on the noncommutative space, and some do not
depend on $\theta$.
(``Topological" is used in a slightly different sense than 
usual and its definition is given below.)
For example, the Euler number of a noncommutative torus is independent of 
the noncommutative parameter $\theta$ and it is defined as a topological 
invariant
by the difference of $K_0$ and $K_1$.
Another example is the possibility to define the instanton number 
(the integral of the first Pontrjagin class) as an integer for Moyal space
\cite{sako2,sako3},
and this fact implies that the instanton number has some kind of 
``topological" nature even for the case that the base manifold is 
noncommutative space.
(Here, we call Moyal space 
 noncommutative Euclidian space whose commutation relations 
of the coordinates are given by $[x^{\mu},x^{\nu}]= i \theta^{\mu \nu}$
, where $\theta_{\mu \nu}$ is an anti-symmetric constant matrix.)
The instanton number does not depend on $\theta$, at least for 
Moyal space.
Also the partition functions of CohFT are examples of the such 
``topological" invariants \cite{sako}.

These observations show that ``topological" charges 
defined by noncommutative field theory have a tendency
to be independence of $\theta$. 
Therefore it is natural to expect the existence of a topological class 
less sensitive than $K$-theory but nontrivial.
Here, we define  an ``insensitive topological invariant"
as follows: If noncommutative manifolds A and B
give the same $K$-group, then the topological invariant defined 
on both A and B 
takes the same value. However, the reverse of this statement is not 
always true.
In short, if K-theory does not distinguish A from B,
then the ``insensitive topological invariant" does not classify them. 
To express thess insensitive topology classes we use ``topological" in the 
above sentences.
One may think that such an insensitive topology
is not useful for geometrical classification.
Possibly,  ``topological" in the above sense might not be suitable 
for the instanton number or the partition function of N.C.CohFT, since there 
is some circumstantial evidence, but this is not proved.
Nevertheless, even if they are not ``topological", 
they have an indisputable value from field theoretical point of view,
since it is possible to classify manifolds by global characters
whose equivalent relations are defined by field theories.
In this sense, this classification is similar to the mirror of 
Calabi-Yau manifolds or to duality in a physical sense.
Therefore, one of the aims of this article is to 
investigate the partition functions of some models of N.C.CohFT
as examples of such ``topological" invariants.

As mentioned above,  N.C.CohFT have the property of $\theta$-shift 
invariance and
the proof of $\theta$-shift invariance is based on the
smoothness for $\theta$ \cite{sako,sako1}.
In some cases, at the commutative point ($\theta=0$) theories have
singularities, as we know e.g. from U(1) instantons.
So, we have to keep in mind that there are difficulties to connect 
a noncommutative 
theory to a commutative theory and the smoothness of $\theta$ for the proof
should be checked whenever we consider new models.  

On the other hand, there are interesting phenomena caused by $\theta$-shift.
For examples, when we consider Moyal spaces, derivative terms 
in the action functional become irrelevant in the large $\theta$ limit.
Then the theory is determined by the potential of the action
and the calculation of the partition function sometimes becomes easy.
If we can compare the moduli space topology in the large 
$\theta$ limit with the one for finite $\theta$,
the $\theta$ invariance of the partition function may be checked.
We shall verify this for one model in this article.

Here, we comment on the relation between \cite{sako} and this article.
As an example of N.C.CohFT, a scalar field theory was 
investigated and its partition function was calculated in \cite{sako}.
This model is essentially equivalent to the model that is 
studied in this article.
We found that the partition function was given as the ``Euler number" of 
a moduli space by using the method of the fundamental theorem of 
Morse theory extended to 
the operator space.
This fact implies that the partition function 
is still the sum of the Euler numbers even 
if the base manifold is a noncommutative space.
But it is not enough to verify the equivalence of the above ``Euler number"
and usual the Euler number defined for commutative manifolds,
because we do not know the connection between the usual Euler number 
and the extension of the fundamental theorem of 
Morse theory to the operator formalism, in the sense of local geometry.
The calculation in \cite{sako} is done by choosing some 
representation of Hilbert space caused from noncommutativity,
and choosing a representation can be understood 
as a gauge fixing.
The computation of \cite{sako} lacks of the view point 
of a local differential geometry of moduli space.
On the other hand, when the moduli spaces are defined as spread commutative
manifolds, their Euler number is given by the Chern-Weil theorem, then 
we expect that the partition function
is obtained by the Chern-Weil theorem.
In other words, we will find that the fundamental theorem of Morse theory 
extended to the operator formulation connects to the usual local geometry
or the usual Euler number on commutative space.
It is worth verifying this statement.
In this article, we demonstrate this for one example.

We remark that the  
operator representation of N.C.field theories 
can be interpreted as an infinite dimensional matrix model.
The partition function of N.C.CohFT is determined by the geometry 
of the moduli 
space of the matrix model.
In particular, when the noncommutative space is a Moyal space,
the matrix model does not include the kinetic terms like the IKKT matrix model
in the $\theta \rightarrow \infty$ limit,
because terms with differential operator
in the Lagrangian like kinetic terms become infinitesimal. 
Then, we can calculate the partition functions 
from potential terms for the Moyal space
by using the matrix models.
This relation between N.C.CohFT and matrix models is also 
important to the matrix models, since this relation allows us to 
investigate the topology of their
moduli spaces by the N.C.CohFT.
Additionally, this correspondence is not just for particular cases.
The connection between 
noncommutative cohomological Yang-Mills theory and
the IKKT matrix model is discussed in this article.
One of our aims is to observe 
these relations between the matrix models and N.C.CohFT.\\

The plan of the article is the following:
In the next section N.C.CohFT is reviewed.
We see that the partition function of N.C.CohFT 
is independent of the deformation parameter
of the $*$ product.
In section 3, we introduce a finite size Hermitian matrix model 
(finite matrix model) as a 0 dimensional cohomological field theory
and we calculate its partition function.
This partition function is determined by merely topological information.
In section 4, we construct some models of 
noncommutative cohomological field theory whose moduli
spaces are defined by projection operators.
Projection operators play an important role in the
topology of noncommutative space, because $K_0$ is
obtained by the Grothendieck construction of equivalent classes of
projection operators.
The partition function of one of the models is given by 
the sum of the Euler numbers of moduli space of the projectors spaces.
In particular, using the result of finite matrix model in 
section 3, the partition function of 
the noncommutative cohomological scalar field theory
on the Moyal plane is obtained in section 4.
Independence from noncommutative parameters is also discussed.
The models that contain derivative terms are investigated 
for the finite noncommutative parameter case and 
the large limit case.
We see that the topology of the moduli spaces of both cases are equivalent.
In section 5, a model mirrored by N.C.CohFT in section 4 
is constructed on COMMUTATIVE space and this model gives 
the model in section 4 by large $N$ dimensional reduction.
We see the connection 
between the model and the homotopy 
classification of vector bundles or topological $K$-theory.
Furthermore, from the point of view of $K_0$ we see that our partition 
function 
of N.C.CohFT is ``topological"
for the Moyal plane and noncommutative torus cases.
In section 6, the correspondence between matrix models and N.C.CohFT 
is investigated for the case of N.C.cohomological Yang-Mills theories. 
In the last section, we summarize this article.

\section{Brief Review of N.C.CohFT }

In this section, we give a brief review of 
cohomological field theory (CohFT) and the nature of 
its noncommutative version.
The CohFT is formulated in several ways \cite{TFT} \cite{witten1}
however we only use 
the Mathai-Quillen formalism in this article.

\subsection{Review of Mathai-Quillen Formalism}
Atiyah and Jeffrey gave a very elegant approach to CohFT \cite{Atiyah-Jeffrey}.
The Atiyah-Jeffrey approach is an infinite dimensional generalization 
of the Mathai-Quillen formalism that is Gaussian shaped Thom forms 
\cite{Mathai-Quillen}.
We recall some well known facts here.
Details can be found
in several lecture notes \cite{Labastida-Lozano},
\cite{Blau} and \cite{Cordes-Moore-Ramgoolam}. \\

For simplicity we only consider the finite dimensional case in this subsection.
Let $X$ be an orientable compact finite dimensional manifold.
For a local coordinate $x$ and Grassmann odd variable $\psi$ 
corresponding to $dx$, we introduce the BRS operator
$\hat{\delta}$:
\begin{eqnarray}
\hat{\delta} x_{\mu}= \psi_{\mu} , \ \ 
\hat{\delta} \psi_{\mu} =0.
\end{eqnarray}
Let us consider a vector bundle $V$ with $2n$ dimensional fiber
and Grassman-odd variables $\chi_a$ and 
Grassmann-even variables $H_a$, $a=1,\cdots 2n$.
For these variables, we define the BRS operator
$\hat{\delta}$ transformations:
\begin{eqnarray} 
\hat{\delta} \chi_{a}=  H_{a} ,\ \
\hat{\delta}  H_{a} =0.
\end{eqnarray}
Note that $\hat{\delta}$ is a nilpotent operator.
Using some section $s$ and a connection $A$ of the vector bundle,
the action of the CohFT is defined by the BRS-exact form:
\begin{eqnarray}
 S&=&
 {\hat{\delta} } \left\{ \frac{1}{2} 
 \chi_a ( 2i s^a + A^{ab}_{\mu} \psi^{\mu} \chi_b +H^a ) \right\}
 \nonumber \\
 &=&\frac{1}{2} \left| s^a \right|^2 
-\frac{1}{2} \chi_a \Omega^{ab}_{\mu \nu} \psi^{\mu} \psi^{\nu} \chi_b
-i\nabla_{\mu} s^a(\psi)^{\mu} \chi_a.
\end{eqnarray}
To get the second equality, we integrate out the auxiliary field $H_a$.
The partition function is defined by
\begin{eqnarray}
\label{Z}
Z= \int {\mathcal D}x {\mathcal D}\psi {\mathcal D}\chi {\mathcal D}H
\exp \left( -S \right).
\end{eqnarray}
In the commutative space, the Mathai-Quillen formalism tells us that
the partition function is a sum of Euler numbers
of the vector bundle on the space ${\mathcal M}=\{ s_a^{-1}(0) \} $
with sign. We can see this fact as follows.
We expand the bosonic part $|s^a|^2$ around the zero section $s^a=0$ as
\begin{eqnarray}
|s^a|^2= (\nabla_{\mu} s^a x^{\mu} )^2 +\cdots \ \ .
\end{eqnarray}
In general, the CohFT is invariant under rescaling of the BRS-exact terms,
then the exact expectation value is given by a Gaussian integral.
The Gaussian integral of the bosonic parts gives
\begin{eqnarray}
1/\sqrt{det|\nabla_{\mu} s^a|^2}. \label{bunbo}
\end{eqnarray}
Note that if the connected submanifolds ${\mathcal M}_k $ defined by 
\begin{eqnarray}
\bigcup_k {\mathcal M}_k :=\{ x | s=0 \}, \\
{\mathcal M}_i \cap {\mathcal M}_j = \emptyset 
\ \makebox{for}\ \  i \neq j \nonumber
\end{eqnarray}
 have finite 
dimension, the Gaussian integral is performed over 
$X \backslash  \{ x | s=0 \}$.
The fermionic non-zero mode $\psi,\chi$ integral is 
\begin{eqnarray}
det(\nabla_{\mu} s^a) \label{bunsi}
\end{eqnarray} 
from the fermionic action 
$\nabla_{\mu} s^a(\psi)^{\mu} \chi_a$.
{} From (\ref{bunbo}) and (\ref{bunsi}), the sign $\epsilon_k = \pm$ is given.
Here, the remaining zero modes of $\psi$ are tangent to ${\mathcal M}_k $
and the zero modes of $\chi$ are understood as a section of the vector 
bundle over ${\mathcal M}_k $. Let $\psi_0$ and $\chi_0$ dentote these 
zero-modes
and $V_k$ be the vector bundle over ${\mathcal M}_k $.
Then the remaining integral over ${\mathcal M}_k $ is expressed as 
\begin{eqnarray}
\int_{{\mathcal M}_k} {\mathcal D} \psi_0 {\mathcal D} \chi_0 
e^{ -\frac{1}{2} \chi_{a0} \Omega^{ab}_{\mu \nu} 
\psi^{\mu}_0 \psi^{\nu}_0 \chi_{b0}}
=\chi(V_k ).
\end{eqnarray}
Here $\Omega_{\mu \nu} $ is the curvature. After using Chern-Weil theorem
the right hand side is given by the
Euler number of the vector bundle $V_k$.
Finally we obtain the partition function 
\begin{eqnarray}
Z=\sum_k { \epsilon_k} \chi (V_k). \label{Z}
\end{eqnarray}

The Cohomological field theories are naive extensions of this 
Mathai-Quillen formalism to the infinitesimal dimensional cases.
The transition to the N.C.CohFT is trivially achieved
by going over to operator valued objects everywhere
or by replacing the product by the $*$ product everywhere.


\subsection{ Some Aspects of N.C.CohFT }
In this subsection 
we review some aspects of N.C.CohFT that are investigated in \cite{sako,sako1}.\\

In this article we use both, the $*$ product formulation and 
the operator formulation \cite{Bayen-etc}.
We define the $*$ product of noncommutative deformation by
using the Poisson bracket $\{\ , \}_{\theta}$ as follows
\begin{eqnarray}
\phi_1 * \phi_2 = \phi_1  \phi_2
+\frac{1}{2}\{\phi_1 , \phi_2\}_{\theta}
+ (\makebox{higher order of $\theta$}), \label{*}
\end{eqnarray}
where $\phi_i$ (i=1,2) are sections of vector bundles
whose base manifold is a Poisson manifold.
Note that the Poisson brackets are defined on Poisson manifolds.
The $*$ product is frequently expressed by $\hbar $ expansion and 
this $\hbar$ is distinguished from the symplectic form used for the 
definition of the 
Poisson bracket.
However, we make no distinction between $\hbar$ and the symplectic form
and collectively call them 
noncommutative parameters $\theta$, hereinafter,
for simplicity. The index $\theta$ of $\{\ , \}_{\theta}$ denotes the set of
noncommutative parameters.
For example, we will use Moyal product for ${\mathbb R}^{2n}$ and 
$T^2$ when we perform concreate calculations in section 4.
In these cases, 
the following Poisson brackets are used, 
\begin{eqnarray}
\{\phi_1 , \phi_2\}_{\theta}=\frac{i}{2 } \theta^{\mu \nu}
(\partial_{\mu} \phi_1 \partial_{\nu} \phi_2-
\partial_{\mu} \phi_2 \partial_{\nu} \phi_1),
\end{eqnarray}
where the noncommutative parameter $\theta^{\mu \nu}$ is a constant 
anti-symmetric matrix.
Then the $*$ product, called the ``Moyal product" \cite{Moyal}, 
for ${\mathbb R}^{2}$ or 
$T^2$ is given by
\begin{eqnarray}
\phi_1 * \phi_2 (x) = 
e^{\frac{i}{2}\theta^{\mu \nu}\partial_{\mu} \partial_{\nu}' }
\phi_1(x) \phi_2(x') |_{x=x'}.
\end{eqnarray}
In the following, $*$ is used for both general Poisson manifolds
and ${\mathbb R}^{2}$ or $T^2$.
So, when we consider ${\mathbb R}^{2}$ or $T^2$, 
we write ``Moyal product" , ``Moyal plane" etc. 
to make a distinction from the general $*$ product.
\\

Let us consider the CohFT on some Poisson manifolds deformed 
by the $*$ product. Take the Lagrangian and the partition function as in the
previous subsection but with infinite dimensions.
Naively, replacing $x$, $\chi$ etc. by some fields 
$\phi^i(x)$, $\chi^a(x)$ etc. 
gives the infinite dimensional extension of the Mathai-Quillen formalism.
Since the action functional is defined by an BRS-exact
functional like $\hat{\delta} V$,
its partition function is invariant under any infinitesimal
transformation $\delta'$ which commutes 
(or anti-commutes) with the BRS transformation: 
\begin{eqnarray}
\label{2.13}
\hat{\delta} \delta' &=& \pm \delta' \hat{\delta}, \nonumber\\
\delta' \ Z_{\theta} &=& \int {\mathcal D}\phi {\mathcal D} \psi {\mathcal D}\chi {\mathcal D}H
\ \ \delta' \left( -\int dx^D \hat{\delta} V \right) \ \exp \left( -S_\theta
\right)
\nonumber\\
&=& \pm \int {\mathcal D}\phi {\mathcal D} \psi {\mathcal D}\chi {\mathcal D}H \ \
\hat{\delta} \left( -\int dx^D \delta' V \right) \ 
\exp \left( -S_\theta \right)
=0.
\end{eqnarray}
Let $\delta_{\theta}$ be the infinitesimal deformation 
operator of 
the noncommutative parameter $\theta$ which operates as
\begin{eqnarray}
\delta_{\theta}\ \theta^{\mu \nu} = \delta \theta^{\mu \nu},
\end{eqnarray}
where $\delta \theta^{\mu \nu}$ are 
some infinitesimal anti-symmetric two form
elements.
To express the dependence on $\theta$, we use $\mbox{\Large $*$}_{\theta}$ as 
the $*$ product defined by (\ref{*}) with noncommutative parameter $\theta$
in the following discussion. 
For $\mbox{\Large $*$}_{\theta}$, the $\delta_{\theta}$ 
operation is represented as 
\begin{eqnarray}
\delta_{\theta}\ \mbox{\Large $*$}_{\theta} = \mbox{\Large $*$}_{\theta+\delta\theta} -\mbox{\Large $*$}_{\theta}.
\end{eqnarray}
Then we see that $\hat{\delta}$ commute with $\delta_{\theta}$
 as follows,
\begin{eqnarray}
\hat{\delta} \delta_{\theta} ( \phi_1 \mbox{\Large $*$}_{\theta} \phi_2 )
&=& \hat{\delta} (\phi_1 \mbox{\Large $*$}_{\theta+\delta \theta} \phi_2 -
\phi_1 \mbox{\Large $*$}_{\theta} \phi_2 )
\nonumber \\
&=& ( \psi_1 \mbox{\Large $*$}_{\theta+\delta \theta} \phi_2 +
(-1)^{P_{\phi_1}}  \phi_1 \mbox{\Large $*$}_{\theta+\delta \theta} \psi_2 )-
\nonumber \\
&&-( \psi_1 \mbox{\Large $*$}_{\theta} \phi_2 +
(-1)^{P_{\phi_1}}  \phi_1 \mbox{\Large $*$}_{\theta} \psi_2 )
\nonumber \\
&=& \delta_{\theta} (\psi_1 \mbox{\Large $*$}_{\theta} \phi_2 +(-1)^{P_{\phi_1}} \phi_1 \mbox{\Large $*$}_{\theta} \psi_2) 
\nonumber \\
&=& \delta_{\theta} \hat{\delta}( \phi_1 \mbox{\Large $*$}_{\theta} \phi_2 ),
\end{eqnarray}
where $\psi_i = \hat{\delta} \phi_i$ and 
$P_{\phi_i}$ is the parity of $\phi_i$. 
This fact shows that the partition function of the N.C.CohFT is invariant 
under the $\theta$ deformation.\\

Note that this proof of invariance under the $\theta$ deformation
is available when the definition of the 
BRS transformation does not include the $*$ product explicitly.
If the $*$ product is used in the definition of 
the BRS transformation, 
the $\delta_{\theta}$ and $\hat{\delta}$ do not
commute, i.e. we can not apply the above proof.
Cohomological Yang-Mills theory is one of such theories,
and this issue will be disscussed in section 6.\\

If we restrict the models to Moyal spaces,
more concrete and interesting properties appear from $\theta$ shifting.  
To clarify their character,
we introduce the rescaling operator $\delta_s$ that satisfies
\begin{eqnarray}
\label{2.4}
    {x'}^{\mu}
&=&
    x^{\mu} -\delta_s x^{\mu}, \\
        \delta_s x^{\mu}
&=&
    (\frac{1}{2}
     \delta \theta^{\mu \nu} (\theta^{-1})_{\nu \rho} )x^{\rho}
\end{eqnarray}
and
\begin{eqnarray}
\label{2.3}
   (1-\delta_s)
   [x^{\mu},x^{\nu}]
        = [{x'}^{\mu},{x'}^{\nu}]
 =
    i({\theta^{\mu \nu}
    -\delta \theta^{\mu \nu}}).
\end{eqnarray}
The transformation matrix is given as
\begin{eqnarray}
\label{2.7}
    J^{\mu}_\rho \equiv {\delta^{\mu}}_{\rho}+
\frac{1}{2} \delta \theta^{\mu \nu} (\theta^{-1})_{\nu \rho},
\end{eqnarray}
and the integral measure is expressed as
\begin{eqnarray}
\label{2.8}
dx^D = \det {\mathbf{J}} {dx'}^{D}, \ \ \ \ \
\frac{\partial}{\partial x^{\mu}} =
(J^{-1} )_{\mu \nu} \frac{\partial}{{\partial x'}^{\nu},},
\end{eqnarray}
where $ \det {\mathbf{J}}$ is the Jacobian. \\
Using these new variables the Moyal product is rewritten as
\begin{eqnarray}
\label{2.9}
(1-\delta_s) (\mbox{\Large $*$}_{\theta}) =
\delta_s (\exp(\frac{i}{2} \overleftarrow{\partial}_{\mu}
(\theta-\delta \theta)^{\mu \nu} \overrightarrow{\partial}_{\nu}) )=
\mbox{\Large $*$}_{\theta -\delta \theta}.
\end{eqnarray}
The above processes are simply changing variables, so 
the theory itself is not changed.
The action before and after this variable change is written as follows.
\begin{eqnarray}
\label{2.10}
S_{\theta}&=& \int dx^D {\mathcal L}(\mbox{\Large $*$}_{\theta}, \partial_{\mu} )
\nonumber\\
          &=&    \int \det {\mathbf{J}} {dx'}^D
              {\mathcal L}(\mbox{\Large $*$}_{\theta-\delta \theta},
              (J^{-1} )^{\mu \nu} \frac{\partial}{\partial {x'}^{\nu}}),
\end{eqnarray}
where ${\mathcal L}(\mbox{\Large $*$}_{\theta}, \partial_{\mu} )$ is an explicit
description to emphasise that the product
of fields is the Moyal product and the Lagrangian contains
derivative terms.

As the next step,
we shift the noncommutative parameter $\theta$ as follows
\begin{eqnarray}
\label{2.11}
\theta \to \theta' =\theta + \delta \theta .
\end{eqnarray}
This deformation changes theories in general.
However, the partition function of the N.C.CohFT 
does not change under this shift as we have seen. 
After changing of variables (\ref{2.4}) and deforming
$\theta$ (\ref{2.11}), the action is expressed as follows. 
\begin{eqnarray}
\label{2.12}
S_{\theta'}=\int \det {\mathbf{J}} {dx'}^D
              {\mathcal L}( \mbox{\Large $*$}_{\theta},
              (J^{-1} )^{\mu \nu} \frac{\partial}{{\partial x'}^{\nu}}).
\end{eqnarray}
Here ${\mathcal L}(\mbox{\Large $*$}_{\theta},
(J^{-1} )^{\mu \nu} \frac{\partial}{{\partial x'}^{\nu}})$
is a Lagrangian in which the multiplication of fields is defined 
by $\mbox{\Large $*$}_{\theta}$ and all differential operators
$\frac{\partial}{{\partial x}^{\mu}}$ in the original
Lagrangian are replaced by 
$(J^{-1} )^{\mu \nu} \frac{\partial}{{\partial x'}^{\nu}}$ without 
derivations in $\mbox{\Large $*$}_{\theta}$. 
{}This action (\ref{2.12}) shows that the $\theta$ deformation is equivalent to
rescaling of $x$ by $\delta_s$, but 
the Moyal product $\mbox{\Large $*$}_{\theta}$ is fixed.
Note that the 
$\theta \rightarrow \infty $ limit is given by omitting
kinetic terms in the action, because the limit 
$\theta^{\mu \nu}
\rightarrow \infty $ means $\det{\bf J}\rightarrow
\infty$ in Eq.(\ref{2.12}) (see also \cite{GMS} and \cite{G-H-S}).
Using this property we investigate both, the large $\theta$ limit case
and the finite $\theta$ case for some N.C.CohFT model 
on the Moyal plane in section 4.

%
\section{Finite Matrix model with Connections}

In this subsection, we study a matrix model and its partition function.
Finite size or infinite size Hermitian matrix models are important 
in physics, even for one-matrix models
(see for example \cite{Alexanrov}, \cite{Nishimura1} and \cite{Nishimura2}).
The model considered here is different from these models, however 
the methods of the analysis done here is also applicable to them 
when we study the geometry of their moduli spaces.
The matrix model of this section is regarded
as an operator representation of 
the N.C. cohomological scalar model of section 4, with 
a cut-off taken in the Hilbert space.
{}From this fact, the calculations performed in this section
make it possible to determine the partition function of the N.C.CohFT
on the Moyal plane in section 4.
(This model is also obtained by a reduction to dim=0 of the model 
in section 5.)
\\

Let $M$ be the set of all
$N\times N$ Hermitian matrices, then this is a $N^2\dim$ Euclidian manifold
${\mathbb R}^{N^2}$.
Let $V$ be a rank $N^2$ (trivial) vector bundle over $M$.
Let $s$ : $M \rightarrow V$ denote
some given section of a trivial bundle. 
We adopt the Killing form as a positive-definite inner product.

We construct the finite matrix model as the 0 dimensional CohFT.
Take some orthonormal basis of $N\times N$ Hermitian matrices 
as a canonical coordinate of $M$, and write
$\phi=(\phi^{ab}) \in M$. 
The other fields (matrices) are introduced 
by the method of general CohFT.
$H^{ab}$ 
is a bosonic auxiliary 
field that is a $N\times N$ Hermitian matrix.
Fermionic matrices are $\psi^{ab}$ and $\chi^{ab}$, i.e. the
BRS partners of $\phi $ and $H$, and
they are $N\times N$ Hermitian matrices, too.
Their BRS transformation is given as 
\begin{eqnarray}
 &&\hat{\delta} \phi = \psi ,\; \hat{\delta} \psi=0 , \;
  \hat{\delta} \chi = H ,\; \hat{\delta} H =0 .
\end{eqnarray}

Let $\nabla$ be a connection 
$\Gamma(V) \rightarrow \Gamma(T^* M \otimes V)=V$,
where $\Gamma(V)$ is a set of all sections. 
Let $A^{{}\ \ kl}_{ji;mn}(\phi)$ be a component of a connection 
1-form in the vector bundle $V$, and 
$e_{ij}$ be a component of a local frame field 
of $V$. Using $e_{ij}$, the relation between $A$ and $\nabla$ is
written as
$\nabla_{ij} e_{kl} = \sum A_{ij;kl}^{{}\ \ mn} e_{mn}$.
In the following, we take the section of the trivial bundle 
as $s(\phi)=\phi(1-\phi)$.
Then the CohFT action is given by
\begin{eqnarray}
 S&=& \sum_{i,j} \hat{\delta}  \{
 \chi^{ij}(2[\phi(1-\phi )]_{ji}
+i \sum_{m,n,k,l} \chi^{mn}A^{{}\ \ kl}_{ji,mn}(\phi) \psi_{kl}  
 -iH_{ij})
 \}.
\end{eqnarray}
After performing the Gaussian integral of $H_{ij}$, the bosonic part of the action becomes
\begin{eqnarray}
\tr (\phi(1-\phi))^2,
\end{eqnarray}
and the fermionic part of the action is
\begin{eqnarray}
 \mathcal{L}_F &=& \tr i \chi  \Big\{
        2(\psi (1-\phi) -\phi \psi ) -
       \sum_{ijklmn} \psi_{ij}  
        \psi_{kl} F(ij,kl ; ab ,mn)  \chi_{mn} 
        \Big\} .
\end{eqnarray}
Here $F(ij,kl ; ab ,mn)$ is the curvature defined by
\begin{eqnarray}
{}&&F(ij,kl ; ab ,mn)\equiv \\
{}&&\frac{\delta}{\delta \phi_{ij}} A^{{}\ \ mn}_{kl;ab}-
\frac{\delta}{\delta \phi_{kl}} A^{{}\ \ mn}_{ij;ab}
+ i \sum_{(c,d)}[A^{{}\ \ cd}_{ij;ab}A^{{}\ \ mn}_{kl;cd}-
 A^{{}\ \ cd}_{kl;ab}A^{{}\ \ mn}_{ij;cd}] \nonumber
\end{eqnarray}
The fixed points of this action are determined by
\begin{eqnarray}
(\phi(1-\phi))=0 .
\end{eqnarray}
Non-zero solutions of $\phi$ are the projection operators $P$
defined by $P^2=P$.
We denote by $P_k$ the projector that restricts a rank $N$ vector space 
to a dimension $k$ vector space.
The set of all $P_k$ is connected and 
\begin{eqnarray}
{\mathcal M}_{k,N} \equiv \{ P_k\} = G_k(N) ,
\end{eqnarray}
where $G_k(N)$ is a Grassman manifold ($\frac{U(N)}{U(k)U(N-k)}$)
whose dimension is $2k(N-k)$.\\


Let us investigate the ${\mathcal M}_k$ from a local geometric aspect.
At first, we prove the non-degeneracy of $s$ in the normal
directions to ${\mathcal M}_k$.
The definition of non-degeneracy is as follows.
Locally one can pick a coordinate $e_{ij}$ (
number of combination $(i,j)$ is $N^2-2k(N-k)$ )
in the directions normal to  
${\mathcal M}_k$  
and a trivialization of $V$ such that
\begin{eqnarray}
s^{ab}=\sum_{i,j}f^{ab}_{ij} e^{ij} , \ &\makebox{ for} &\ 
(i,j) , (a,b) \in {\bf N}
\label{normal}\\
s^{ab}=0 \ , \ &\makebox{ for} &  (a,b) \in {\bf T}.
\end{eqnarray}
Here, ${\bf N}$ and ${\bf T}$ are sets of indices $(i,j)$
and the numbers of their elements are $N^2-2k(N-k)$ and  $2k(N-k) $, 
respectively.
Let us prove this non-degeneracy of ${\mathcal M}_k$.
After an appropriate coordinate choice,
we can take a rank $k$ solution $P_k \in {\mathcal M}_k$ as 
\begin{eqnarray}
P_k=  
\left(
\begin{array}{c|c}
{\bf 1}_k & 0 \\
\hline
0&0
\end{array}
\right),
\end{eqnarray}
where $P$ is a $N\times N$ matrix valued projection operator
and ${\bf 1}_k $ is the $k\times k$ unit matrix.
The (co)tangent vectors at this point are determined by
variation of $\phi$ equation around this solution;
\begin{eqnarray}
\delta \phi ( 1-P_k ) - P_k \delta \phi=0.
\end{eqnarray}
Its solutions are given by
\begin{eqnarray}
\delta \phi_{ij}=0, \  \delta \phi_{mn}=0, \ 
\delta \phi_{in} = \delta \bar{\phi}_{ni},\ \ 
&\makebox{for}&\  i,j \in \{1,2, \cdots ,k \} \  
\nonumber \\
&&
  m, n \in\{ k+1, \cdots ,N \}.
\end{eqnarray}
Here $\bar{\phi}$ is the complex conjugate of $\phi$.
We can chose a 
$2(N-k)k\dim$ orthonormal basis of $N\times N$ matrices $\delta \phi$:
\begin{eqnarray}
(\phi^{(in)}_R)=
\left(
\begin{array}{c|c}
O & (\delta_{in}) \\
\hline
(\delta_{ni}) &O
\end{array}
\right) \ , \ 
(\phi^{(in)}_I)=
\left(
\begin{array}{c|c}
O & i(\delta_{in}) \\
\hline
-i(\delta_{ni}) &O
\end{array}
\right)
\end{eqnarray}
where $i \in \{1,2, \cdots ,k \}$ and 
$n \in\{ k+1, \cdots ,N \}$.
Let us not confuse ``$i$" of $\sqrt{-1}$ and index in this article.
On the other hand, it is possible to choose a basis of the normal 
direction $e_{normal}$ as a Lie algebra of 
$U(k)\times U(N-k)$ whose non-zero
elements lie only in the block diagonal part i.e. $(e_{normal})_{in}=0$ for 
$i \in \{1,2, \cdots ,k \}$ and 
$n \in\{ k+1, \cdots ,N \}$. 
(Note that $\tr \phi^{(in)}_I e_{normal} = \tr \phi^{(in)}_R e_{normal} =0$
shows that the direction of $e_{normal}$ is normal to $\delta \phi$.)
This basis of the Lie algebra may be chosen to be non-degenerate.
For example,  
we can take this non-degenerate $N^2-2k(N-k)\dim$ 
basis $e_{normal}$ as,
\begin{eqnarray}
(e^{(ij)}_{normal}) 
= \left\{ 
\begin{array}{c}
\left(
\begin{array}{c|c}
U^k_{i,j} & O \\
\hline
O & O
\end{array}
\right)\ \ , \  \makebox{for} \ i \ \makebox{and}\ j \in \{ 1, \cdots ,k\}\\
\makebox{} \\
\left(
\begin{array}{c|c}
O & O \\
\hline
O & U^{N-k}_{i,j}
\end{array}
\right)\ \ , \  i \ \makebox{and}\  j  \in \{ k+1, \cdots ,N \}
\end{array}
\right. 
\end{eqnarray}
 where $\{U^{N-k}_{i,j;a,b} \}$ is a orthonormal basis of 
$u(k)$  and $\{U^{N-k}_{i,j;a,b} \}$ is one of $u(N-k)$.
We found the local coordinate $e_{normal}$ in the directions
normal to ${\mathcal M}_k$ such that (\ref{normal}) holds.
This shows non-degeneracy.
This discussion for non-degeneracy is parallel to the one in 
\cite{Vafa-Witten}.

Let us investigate the mass matrix of fermions near the ${\mathcal M}_k$
and the fermionic zero-modes.
The $\chi$ equation and the $\psi$
equation are
\begin{eqnarray}
\psi (1-P) -P \psi =0 , \ 
\makebox{and} \ \chi (1-P) -P \chi=0, 
\end{eqnarray}
where we neglect nonlinear terms.
Note that $f^{ab}_{ij} $ in (\ref{normal}) is the mass matrix of $\chi$ and $\psi$
near ${\mathcal M}_k$.
Using the $\chi $ equation, we see that the 
massless components of 
$\psi$ are those ones that are tangent to ${\mathcal M}_k$.
There are massless components of $\chi^{ab}$ that are
regarded as the above trivialization i.e. $(a,b) \in {\bf T}$.
Furthermore we can understand from the $\psi$ equation
that the $\chi $ zero-modes are sections of the (co)tangent 
bundle of ${\mathcal M}_{k,N}$.

Now we evaluate the integral for $Z$.
The mass components integral gives overall factor 
$(-1)^{k^2}=(-1)^k$ ( see \cite{sako} ).
Recall that the moduli space
$\{\phi |s=0 \} = \bigcup_{k} \{ P_k\} $
and $\{ P_k\} =G_k(N)$.
The Poincare polynomial of the Grassman manifold
is given as 
\begin{eqnarray}
P_t(G_k(N))=
\frac{(1-t^2)\cdots (1-t^{2N})}{(1-t^2)\cdots (1-t^{2(N-k)})(1-t^2)\cdots
(1-t^{2k})}. \nonumber
\end{eqnarray}
(See for example \cite{Bott}.)
Using these results and (\ref{Z}), the partition function is written as 
\begin{eqnarray}
Z=  \sum_{k=0}^{N} (-1)^k P_{-1}(G_k(N)). \label{z_matrix}
\end{eqnarray}
When we take $t=\pm 1$,
the Poincare polynomial becomes the number of combinations,  
\begin{eqnarray}
P_{\pm 1}(G_k(N))= \frac{N!}{k!(N-k)!} \equiv \label{euler_grass}
{N \atopwithdelims() k} \ .
\label{p_1}
\end{eqnarray}
The proof of (\ref{p_1}) is given as follows.
\begin{eqnarray}
P_{\pm 1} (G_k(N))&=& \left.
\frac{(1-t^2)\cdots (1-t^{2N})}{(1-t^2)\cdots (1-t^{2(N-k)})(1-t^2)\cdots (1-t^{2k})} \right|_{t=1} \nonumber \\
&=& \left.
\frac{(1-t^{2(N-k+1)})\cdots (1-t^{2N})}{(1-t^2)\cdots (1-t^{2k})} \right|_{t=1}.
\end{eqnarray}
After replacing $t^2$ by a positive real number $x$,
\begin{eqnarray}
&&P_{\pm 1} (G_k(N))
= \left.
\frac{(1-x^{(N-k+1)})\cdots (1-x^{N})}{(1-x)\cdots (1-x^{k})} \right|_{x=1}
\nonumber \\
&&= \left.
\frac{\{(1-x)(1+x+\cdots +x^{N-k})\}\cdots \{(1-x)(1+x+\cdots +x^{N-1})\}}{\{(1-x)\}\{(1-x)(1+x)\}\cdots \{(1-x)(1+x+\cdots +x^{k-1})\}} \right|_{x=1}
\nonumber \\
&&= \frac{(N-k+1)(N-k+2)\cdots N}{1\cdot 2 \cdots k}
= {N \atopwithdelims() k}.
\end{eqnarray}
This is what we were aiming for. 
{}From (\ref{z_matrix}), (\ref{euler_grass}) and the binomial theorem,
the final result is then
\begin{eqnarray}
Z=  \sum_{k=0}^{N} (-1)^k 1^{N-k} P_{-1}(G_k(N)) = (1-1)^N =0 .
\end{eqnarray}

The calculation of the finite matrix model in this section
will be used directly in the noncommutative 
cohomological scalar model in the 
next section.


\section{N.C.Cohomological Scalar Model} 

In this section, we study some N.C.cohomological scalar models
and evaluate their partition functions for Moyal space
by using the matrix model partition function of the 
previous section.
We also check the $\theta$-shift invariance of $Z$.


\subsection{N.C cohomological scalar model }

Let $M$ be a $2n$ dimensional Poisson manifold with Riemannian metric.
Let $\phi$ and $H$ be real scalar fields on $M$ and, 
$\psi$ and $\chi$ be the BRS partner fermionic 
scalar fields of $\phi$ and $H$, respectively.
In other words, $(\phi , H , \psi , \chi )$
are elements of $\Omega^0(M)$ with ghost number
$(0,0,1,-1)$ and parity (even, even, odd, odd).

We introduce a nilpotent operator $\hat{\delta}$, i.e.
\begin{equation}
 \hat{\delta}^2 =0  ,
\end{equation}
as a BRS operator whose transformation is given by
\begin{eqnarray}
 \hat{\delta} \phi = \psi ,\
 \hat{\delta} \chi = H ,\ \hat{\delta} \psi 
  = \hat{\delta} H  =0 . \label{BRS}
\end{eqnarray}

We consider the deformation quantization defined 
by some $*$ product.
($*$ product exist on arbitrary Poisson manifolds
\cite{Kontsevich}.)

We consider two actions :
\begin{eqnarray}
 S_1&=&\int_M dx^D \sqrt{g}\mathcal{L}  \label{S_1}\\
 S_2 &=& S_1 + S_{top},   \label{S_2}
\end{eqnarray}
where the Lagrangian $\mathcal{L}$ is given by
\begin{eqnarray} \label{Lag}
\mathcal{L}&=&
 {\hat{\delta} }\left( \frac{1}{2} 
 \chi * \left( 2(\phi *( 1-\phi ) ) + 
 \frac{2i}{g} \int d^{2n} z d^{2n} y \psi (z) A(z ; x ,y)  \chi(y) 
 -iH \right) \right).
 \nonumber \\
 \label{Lag}
\end{eqnarray}
Here, $g$ is a coupling constant, 
$x, y, z \in M$, and $A(z ; x ,y) $ is some functional of 
$\phi$ that should be defined as a 
connection on the trivial bundle over the set of all $\phi$. 
$A(z ; x ,y) $ is an anti-symmetric matrix with respect to $x$ and $y$, and 
the multiplication between $A(z ; x ,y), \psi (z)$ and $ \chi(y)$  is 
not $*$ multiplication since the trace operation (integral) over $z$ and 
$y$ has been performed.
(However, we can also express their products by $*$ product in the integral.)
This may look like some strange kind of non-local interaction,
but it is possible to regard this as an integral kernel.
In many cases, deformation quantization is introduced by an integral
kernel, therefore such kind of non-local interactions are 
not unusual in noncommutative field theory.
The precise definition of $A(z ; x ,y)$ depends on $M$ and 
the deformation by $*$, the connection is formally introduced, here.
When we consider the ${\mathbb R}^2$
case in the following subsection, we will verify 
that $A(z ; x ,y) $ is a connection 
and particularly that it becomes a nontrivial connection on a 
submanifold of $\{ \phi \}$.
Especially in conjunction with the matrix model in previous section,
after using the Weyl correspondence, we can regard $A(z ; x ,y) $
as the usual connection of the (co)tangent vector bundle over some 
Grassman manifold 
that appears as a moduli space of $\phi$.\\
The topological action in $S_2$ is 
\begin{eqnarray}
S_{top}= g' \tau_{2n}({\mathcal F},\cdots ,{\mathcal F}),
\end{eqnarray}
where $g'$ is a coupling constant and ${\mathcal F}$ is defined by
\begin{eqnarray}
{\mathcal F}_{ij}= [ \phi \partial_i \phi , \phi \partial_j \phi ].
\end{eqnarray}
This action itself is not topological, but in our case
$\phi$ is replaced by projection operators.
In such a case, we can regard $S_{top}$ as Connes's Chern character.
Connes's Chern character homomorphism is:
\begin{eqnarray}
ch_{2n} &:& K_0({\mathcal A}) \rightarrow HC_{2n} ({\mathcal A}) \nonumber \\
ch_{2n}(p)&=& \sum_{n=0}^{\infty} \tau_{2n}({ f},\cdots ,{ f})
\end{eqnarray}
where $f_{ij}= [p \partial_i p , p \partial_j p] $. 
It is worth emphasizing that $S_{top}$ is not invariant in general 
under changing the noncommutative parameter $\theta$, 
since it is not a BRS exact action.
Indeed $ch_{2n}(p)$ apparently depends on $\theta$ 
for noncommutative torus example.
Therefore $S_2$ is not suitable if we
are interested in only constructing the 
$\theta$-shift invariant theory.
However, there is another motivation to construct the
N.C.CohFT, that is to construct some ``topological" invariant.
In the commutative case, we often add a topological action to the 
BRS exact one, and the topological terms play important roles.
In analogy with commutative CohFT, 
it seems useful to consider both the $S_1$ and the $S_2$ case.


The Lagrangian $\mathcal{L}$ without the $S_{top}$ part 
is divided into a bosonic part $\mathcal{L}_B$ 
and fermionic part $\mathcal{L}_F$:  
\begin{eqnarray}
 \mathcal{L}&=&\mathcal{L}_B+\mathcal{L}_F , \\
 \mathcal{L}_B &=& |\phi * (1-\phi)  |^2 \ ,  \label{LB}
 \\ 
 \mathcal{L}_F &=& i\chi * \Big\{
        2(\psi *(1-\phi) - \phi *\psi )
       \nonumber \\
        && {}\hspace{8mm} - {}\frac{i}{2g}\int d^n z d^n w d^n y \psi (z) 
        \psi(w) F(z,w ; x ,y)  \chi(y) 
        \Big\}. \nonumber
\end{eqnarray}
Here  $F(z,w ; x ,y)$ is defined by
\begin{eqnarray}
\frac{ \delta A(z ; x ,y)}{\delta \phi(w) } &-&
\frac{ \delta A(w ; x ,y)}{\delta \phi(z) }+  \\
&+& \frac{i}{g}\int d^n u  \big( \ A(z ; x,u) A(w; u,y)
-  \ A(w ; x,u) A(z; u,y) \big), \nonumber
\end{eqnarray}
and it corresponds to the curvature.\\

{}From the general argument of the Mathai-Quillen formalism
and the similar analysis as in the previous section, 
the partition function of this theory
is given by the sum of the Euler numbers of the solution space of $\phi$.
{}From Eq.(\ref{LB}), the fixed point loci of $\phi$ 
are given by the set of all projection operators $P$,
i.e. $P * P = P$, and they are called GMS solitons \cite{GMS}. 
We denote by ${\mathcal M}_k $ the set of projections 
distinguished by index $k$. 
An example of the index $k$ is the rank of projections for the case that we 
can define the 
rank by a discrete number.
If there is ghost number anomaly, the partition function vanishes in general.
In our case, there is no ghost number anomaly as we saw in  
section 3, thus we get some nontrivial partition functions 
for $S_1$ and $S_2$:
\begin{eqnarray}
Z_1 &=& \sum_k \epsilon_k \chi({\mathcal M}_k ), \\
Z_2 &=& \sum_k \epsilon_k \chi({\mathcal M}_k ) e^{g'\tau_{2n}(k)}
\end{eqnarray}
where $\chi({\mathcal M}_k )$ is the Euler number of ${\mathcal M}_k $ and 
$\epsilon_k $ gives a sign $\pm$.

When we consider the noncommutative theory from the topological view point, 
the most important operators are projectors and unitary operators
since they define $K_0$ and $K_1$.
This partition function is a sum of integer valued Euler numbers 
of the sets of all projections 
that construct the $K_0$ elements when the moduli space is a manifold.
So it is natural to expect the partition function to be ``topological".

Concrete calculation of the partition function will be done for the 
Moyal plane case below. We are interested in whether 
the ``topological" quantity is invariant under the continuous change 
of the noncommutative parameter.
For the $S_1$ model of this section, it is clear that 
the partition function is invariant under the $\theta$ change
as far as there is no singularity.
A more interesting case is when the Lagrangian has kinetic terms.
To investigate the behavior of the partition function whose Lagrangian
contains kinetic terms, we slightly deform our models 
in the following.


\subsection{N.C. cohomological scalar model with kinetic terms}

Let $M$ be a $2n$ dimensional Poisson manifold with a Riemannian metric.
Let $\phi$ and $H$ be real scalar fields on $M$ and, 
$\psi$ and $\chi$ be $\phi$ and $H$'s BRS partner fermionic 
scalar fields.
Let 
$B_{\mu}$ and $H_{\mu}$ be real vector fields 
and  $\psi_{\mu}$ and $\chi_{\mu}$ be BRS partner 
fermionic vector fields of $B_{\mu}$ and $H_{\mu}$.
In other words, $(\phi , H , \psi , \chi )$
are elements of $\Omega^0(M)$ with ghost number
$(0,0,1,-1)$ and parity (even, even, odd, odd).
$(B_{\mu} , H_{\mu} , \psi_{\mu} , \chi_{\mu} )$
are elements of $\Omega^1(M)$ with ghost number
$(0,0,1,-1)$ and parity (even, even, odd, odd).
The BRS operator transformation is given by
\begin{eqnarray}
 \hat{\delta} \phi = \psi &,&
 \hat{\delta} \chi = H ,\ \hat{\delta} \psi 
  = \hat{\delta} H  =0 ,\
 \hat{\delta} B^{\mu} = \psi^{\mu} ,\nonumber \\ 
&& {}\   \hat{\delta} \chi^{\mu} = H^{\mu} ,\ 
  \hat{\delta} \psi^{\mu} 
  = \hat{\delta} H^{\mu} = 0 .
\end{eqnarray}

One of our interests is to investigate the behavior
of the partition function of N.C.CohFT under a 
change of the noncommutative parameter.
It is difficult to study the general case of deformation quantization.
Therefore, we put an assumption  
in this subsection such that 
terms including derivatives 
like kinetic terms become irrelevant in the 
large noncommutative parameter limit ($\theta \rightarrow \infty$ )
as far as evaluating perturbative contribution is concerned.
For example, when we consider the deformation of ${\mathbb R}^d$
by the Moyal product, only the potential terms become relevant
 in the $\theta \rightarrow \infty$ limit \cite{sako,GMS}.
Note that we make this assumption only for 
simplicity the of calculation.
The invariance under a change of $\theta$ is essential for us, 
and this is not affected by our assumption.
\\

Similar to the previous subsection, 
we consider two types of action:
\begin{eqnarray}
 S_1&=&\int_M dx^D \sqrt{g}\mathcal{L}  \label{S_1withKinetic} \\
 S_2 &=& S_1 + S_{top},  \label{S_2withKinetic}
\end{eqnarray}
where the Lagrangian is slightly different from (\ref{Lag}),
\begin{eqnarray}
&&\mathcal{L}= \nonumber \\
&& {\hat{\delta} }\left( 
 \chi * \left( (\phi *( 1-\phi ) -i\partial_{\mu} B^{\mu}) + 
 \frac{i}{g} \int d^n z d^n y \psi (z) A(z ; x ,y)  \chi(y) 
 -iH \right) \right)
 \nonumber \\
 &&
 +\hat{\delta}  \left(\frac{1}{2}
    \chi^{\mu} *( 2(i\partial_{\mu} \phi + B_{\mu} )- iH_{\mu})\right).
\end{eqnarray}

As noted in the previous subsection, 
the topological term $S_{top}$ has noncommutative parameter $\theta$
dependence in general. 
One such example is the noncommutative torus.
On the other hand, the Moyal plane theory
does not depend on $\theta$.
When we construct a $\theta$ independent ``topological" invariant $Z$,
we find whether we can add $S_{top}$
to the action $S_1$ obtained from the K-theory (cyclic cohomology) 
information of the base manifold.\\

The Lagrangian $\mathcal{L}$ without $S_{top}$ term 
is divided into bosonic part and fermionic part:  
\begin{eqnarray}
 \mathcal{L}&=&\mathcal{L}_B+\mathcal{L}_F , \\
 \mathcal{L}_B &=& |\phi * (1-\phi)  -i\partial_{\mu} B^{\mu}|^2 +
 |i\partial_{\mu} \phi + B_{\mu} |^2 ,\label{boson}\\
 \mathcal{L}_F &=& i\chi * \Big\{
        2(\psi *(1-\phi) - \phi *\psi 
        -i\partial_{\mu}\psi^{\mu})- \label{fermion}\\
      &&\!\!\!\!\!\!\!\!\!\!\!\!\!\!
        - \frac{i}{2g}\int d^n z d^n w d^n y \psi (z) 
        \psi(w) F(z,w ; x ,y)  \chi(y) 
        \Big\}
    +i\chi^{\mu}* \left\{ 2i\partial_{\mu}\psi+2\psi_{\mu}\right\}.
    \nonumber
\end{eqnarray}

Note that this theory is invariant under arbitrary 
$A$ deformation ($ A \rightarrow A + \delta A$)
and deformation of the coupling constant $g$.
In the following subsections, 
we investigate the deformation of 
moduli space and the invariance of
partition function
under change of $\theta$. 
If we observe $\theta \rightarrow \infty$ in the Moyal plane case
by using the scaling method discussed in section 2, the 
$F(z,w ; x ,y)$ contribution to the partition function becomes 
dominating since each integral measure $d^2 z , d^2 w$ and $d^2 y$ 
is of order $\theta$. Then, the surviving terms in the limit are not
the BRS exact terms.
Therefore, we have to tune the other parameters in such a limit 
if we want to use the convenient methods of CohFT.
{}From the fact that the partition function has symmetry
under arbitrary 
$g$ and $A$ variation, we can fit $g$ without changing $Z$
for surviving terms which are BRS exact terms in the limit $\theta \rightarrow \infty$.

\subsection{Moyal plane case in $\theta\rightarrow\infty$}
In this subsection, 
the partition function of the 
N.C.cohomological scalar model is calculated. 
To compute it explicitly, we consider the
two dimension Moyal plane.
There are two reasons for choosing the Moyal plane here. 
First, the Moyal plane satisfies the 
assumption of the previous subsection that derivative terms like kinetic terms 
in the Lagrangian become irrelevant in  
$\theta \rightarrow \infty$.
The other reason is that the rank of a projection operator is defined 
by an integer. 
{}From this, the solution space of 
$\phi $ is given by a Grassmann manifold the properties of which
are well known.
In particular, if we represent our theory by operators, this theory 
can be regarded as an infinite dimensional
matrix model.
It is possible to represent
noncommutative Euclidian plane
by a Hilbert space and we can chose some set of eigenvectors 
with discrete eigenvalues
as the basis of this Hilbert space, like for example Fock states.
So if we introduce a cut-off in this Hilbert space,
we can regard our model as the finite matrix model which we 
discussed in section 3.\\

%
%
%
%
%
%
%

\newcommand{\hphi}{\hat{\phi}}
\newcommand{\hB}{\hat{B}}
\newcommand{\hpsi}{\hat{\psi}}
\newcommand{\hchi}{\hat{\chi}}
\newcommand{\hH}{\hat{H}}

So far we have used the $*$ product representation of
noncommutative field theory, however,
in this subsection the operator representation is used 
since it is convenient to see the relation between the 
finite matrix model and the large $\theta$ N.C. cohomological
scalar model. 

In $\theta \rightarrow \infty$,
we can ignore the terms including derivative as we saw in section 2.
Then, the remaining part of the action in the operator formalism is 
\begin{eqnarray}
 S_{\infty}&=& 
 \sum_{i,j} \hat{\delta}  \{
 \chi^{ij}(2[\phi(1-\phi )]_{ji}
+i (\sum_{m,n,k,l} \chi^{mn} {A}^{{}\ \ kl}_{ji,mn}(\hphi) \psi_{kl} ) 
-iH_{ij})
 \} \nonumber \\
  &&+Tr \hat{\delta}  \{
 \hchi^{\mu}(2\hB_{\mu}-i\hH_{\mu})
 \}, 
\end{eqnarray}
where $\hphi,\ \hpsi,\ \cdots$ are 
operator representation of $\phi$, $\psi,\ \cdots$ 
that have infinite dimensional matrix representation
$\hphi = \sum_{ij} |i\rangle \phi_{ij} \langle j|$ in some 
complete system $ \{ |i\rangle \}$.

We introduce a cut-off to restrict the Hilbert space to 
a finite $N$ dimension vector space.
Let $\{| i \rangle | i= 1, \cdots ,N \} $ be a set of orthonormal basis.
Using this representation, 
the operators $\hphi$, $\hH$, $\hpsi$, $\cdots$  
are expressed by $N\times N$ Hermitian matrices, i.e.
$\hphi \rightarrow ( \phi_{ij} )$ etc.
After integration of $B_{\mu}, H_{\mu}, \chi_{\mu}$ and $\psi_{\mu}$, 
we find that this model is equivalent to the finite dimensional 
matrix model discussed in section 3.

The bosonic part of the action is
\begin{eqnarray}
Tr \{(\hphi(1-\hphi))^2 + \hB_{\mu} \hB^{\mu} \}.
\end{eqnarray}
The fixed point locus is determined by
$(\hphi(1-\hphi))=0$ and $\hB_{\mu}=0$.
The solution is given by $\phi=P$, where $P$ is an arbitrary projection 
operator,
which is called GMS soliton.
The moduli space is obtained as a set of Grassman manifolds
$\{ G_k(N) \} := \{\frac{U(N)}{U(k)\times U(N-k)} \}$
since the rank $k$ projection operator 
determines a subspace the codimension of which is $N-k$.
This solution of rank $k$ projector is interpreted 
as symmetry breaking from
$U(N)$ to ${U(k)\times U(N-k)}$.

On the other hand, the integration of the fermionic part generates the
Euler numbers of the Grassmann manifolds that are given in section 3. 
For the Moyal plane, the topological term $S_{top}$ ($ch_2$) is $g'k$ 
when the solution of $\hphi$ is given by a rank $k$ projection operator.
Note that the value of $ch_2$ 
is independent of $\theta$ for the Moyal plane (see for example \cite{Matsuo}).

Using the Euler number of the Grassman manifolds and
the contribution from the topological term, the partition function is 
then
\begin{eqnarray}
Z_2&=& \lim_{N\rightarrow \infty} \sum_{k=0}^{N} P_{-1}(G_k(N)) e^{g'k} (-1)^{N-k}
\nonumber \\
&=& \lim_{N\rightarrow \infty}(1-e^{g'})^N , 
\label{z2}
\end{eqnarray}
where we take $N \rightarrow \infty $ after using 
the result from the finite matrix model.

If we take $S_1$ as the total action of the theory, 
the partition function is given by (\ref{z2}) with the condition 
$g'=0$, then
\begin{eqnarray}
Z_1=0.
\end{eqnarray}

It is worth to comment here on the introduction of the cut-off in the 
above analysis.
It is a well known fact that some properties of 
noncommutative field theories are originated in the characteristic 
nature of infinite dimensional Hilbert space.
For example, the trace of a commutation $\tr(AB-BA)$ does not vanish 
in noncommutative theories in general. This phenomenom 
does not exist in the finite matrix model.
So one might think that we have to add some correction in order to 
account for this effect from infinite dimension to the above 
partition functions.
However it turns out that we do not have to 
correct the partition function.
First, we consider the real scalar field $\phi$, its fixed 
point being given by a projector in this case.
If the solution is given by a shift operator like the complex scalar field 
case in \cite{Harvey,Witten-N}, 
then the calculation does not close for the case of finite size 
matrices, even though the trace operation is done. 
On the other hand, our solutions are given by projection operators in 
this case,
thus there is a possibility that the calculation closes in the finite 
Hilbert space.
Additionally, even if we treat the shift operator, 
there is a way to take the infinite dimension effect
into account.
The method is to put a cut-off only for the initial and final states, 
to define the trace operation for finite matrices, i.e. 
the intermediate states are not restricted by the cut-off,
(see \cite{sako2,sako3} for details).
Using such methods we can estimate the 
effect of infinite dimension, like the shift operator, 
by finite size computation.
{}The other reason is that we should discuss the partition function 
in the terms of the weak topology since the trace operation is performed 
in the partition function calculation. 
Thus, it is difficult to distinguish $U({\mathcal H})$ from 
$U(\infty ) = \lim_{N \rightarrow \infty} U(N)$ by our calculation.
{}From these facts, it is reasonable 
to evaluate the partition function by using the finite matrix model.



\subsection{finite $\theta $}
One of our aims is to confirm that the partition function 
does not change  under a change of the noncommutative parameter.
The proof of the invariance under the $\theta$-shift is based on the 
smoothness of $\theta$. So, we have to check the smoothness for each model.
In the previous subsection, we considered the $\theta \rightarrow \infty$ 
case and we calculated the partition function 
of the N.C.cohomological scalar model on the
2-dimensional Moyal space by using the result of the finite matrix model. 
Obeying the general property of N.C.CohFT, for
finite $\theta$, we expect that the partition function takes 
the same value as Eq.(\ref{z2}).
This statement is realized when
the moduli space deforms smoothly and its topology does not change 
under the variation of $\theta$.
Therefore, let us compare the moduli space of the large $\theta$
limit with the one of finite $\theta$ in this subsection.
\\

It is difficult to analyze the arbitrary finite $\theta$ case
because derivative terms and the nonlinear terms are 
intertwining. Therefore our strategy is to analyze the moduli space 
deformation from large $\theta$ limit
perturbatively.
Let $\phi_{0}$ and $B_{\mu 0}$ be large $\theta$ limit solutions of 
$\phi$ and $B_{\mu}$ i.e. $\phi_{0}= P,\ B_{\mu 0}=0$.
We consider that the fields belong to 
$C^{\infty}({\mathbb R}^2)[[ 1/\sqrt{\theta} ]]$.
$\phi$ and $B_{\mu}$ are expanded as
\begin{eqnarray}
\phi=\phi_0 +  \frac{1}{\sqrt{\theta}}\phi_1 +\cdots,\ 
B_{\mu }= B_{\mu 0}+ \frac{1}{\sqrt{\theta}}B_{\mu 1} +\cdots ,
\end{eqnarray}
and we substitute them into the action.
The leading order bosonic action is then
\begin{eqnarray}
\frac{1}{\theta}\tr |\phi_1 (\phi_0-1) + \phi_0 \phi_1|^2
+\frac{1}{\theta} \tr |\partial_{\mu} \phi_0 +B_{\mu 1}|^2 \\
=\frac{1}{\theta} \tr \left\{ |\phi_1 (P-1)|^2 + |P \phi_1|^2
+|\partial_{\mu} P +B_{\mu 1}|^2 \right\} 
\end{eqnarray}

Let $|P,i\rangle $ be an eigenvector of the projector $P$ with eigenvalue $1$,
 i.e. $P |P,i\rangle = |P,i\rangle $.
Using this vector, $\sum_{i,j}|1-P,i\rangle a_{ij} \langle P , j|+ h.c.$
is a solution of $\phi_1$, where $(a_{ij})$ is a Hermitian matrix.
However, the deformation of the moduli space from $ \{ a_{1,ij} \}$ is 
trivial and retractable.
On the other hand, $B_{\mu 1}=-\partial_{\mu} P$.
$B_{\mu }$ is deformed but it is determined completely 
by the given $P$.
Therefore the moduli space topology is not changed at all.
In other words, we can deform the moduli space 
smoothly.
This result is consistent with our expectation, hence the partition function 
is invariant under $\theta$ deformation.


\section{ K-theory and Cohomological Scalar model }

We discuss the relation between our theory and K-theory
in this section.

\subsection{Commutative CohFT and Homotopy of Vector bundle}

The relation between a model of CohFT on a COMMUTATIVE space
and the homotopy of the classifying map of a vector bundle are 
studied in this subsection. 
The model is deeply related to 
the N.C.CohFT models which are treated in section 4.
By means of this model we will find an analogy of 
correspondence between our N.C.cohomological scalar model
 and algebraic K-theory, namely 
the
correspondence between CohFT and topological K-theory.\\

Let 
$M$ be a $n \dim$ Riemannian manifold,
$V$ be a rank $N$ trivial vector bundle.
\begin{eqnarray}
\phi : M &\rightarrow &H  \nonumber \\
       x &\mapsto &\phi^{ab}(x) \in H ,\ a , b \in \{1, \cdots ,N \}
\end{eqnarray}
where $H$ is set of all $N\times N$ Hermitian matrices i.e.
$H \equiv \{ h | h^{ab}=\bar{h}^{ba}\}$.
In other words, $\phi$ is a $N\times N$ Hermitian matrix valued 
scalar field on M.
$N\times N$ Hermitian matrix valued scalar fields 
$\phi^{ab}(x)$ and $H^{ab}(x)$ have ghost number
$0$ and the fermionic BRS partners
$\psi^{ab}(x)$ and $\chi^{ab}(x)$ have ghost number $1$ and $-1$, respectively.
The BRS transformation is similar to the previous one, however 
there is a difference caused by the $U(N)$ gauge symmetry.
\footnote{The theory of this subsection has $U(N)$ gauge symmetry.
Since gauge symmetry is not our main subject here, we do not 
discuss the technical problems related with gauge symmetry.}
The BRS operator is nilpotent up to gauge transformation $\delta_g$, 
i.e.
$\hat{\delta}^2=\delta_g $.
We denote by $c(x)$ the scalar field corresponding to a local gauge 
parameter with ghost number 2, then the explicit BRS transformation is given by
\begin{eqnarray}
 \hat{\delta} \phi(x) = \psi(x) , \ 
 \hat{\delta} \chi(x) = H(x) &,\ & \hat{\delta} c(x) =0 \nonumber \\
  \hat{\delta} \psi(x) 
  =\delta_g \phi(x) =i[c(x) ,\phi(x)] &, \ &  
  \hat{\delta} H(x)  = \delta_g \chi(x) = i[c(x) ,\chi(x)]. 
\end{eqnarray}
We introduce the following action;
\begin{equation}
 S=S_0 + S_p +S_g \label{actionCohFTgauge}
\end{equation}
\begin{eqnarray}
 S_0&=&\int_M tr \hat{\delta}  \{
 \frac{1}{2} \chi(2 \phi (1-\phi)-iH)
 \},  \label{CCoh}
\end{eqnarray}
where $S_0$ has $U(N)$ gauge symmetry and we have to 
project out the pure gauge degrees of freedom.
Therefore, we introduce $S_p$ for the projection to the gauge horizontal part
and $S_g$ for the gauge fixing.

After performing the Gaussian integral, the bosonic part of $S_0$  is 
\begin{eqnarray}
(\phi(1-\phi))^2 ,
\end{eqnarray}
and the fermionic action is 
\begin{eqnarray}
\chi ( -2\psi (1- \phi) + 2\phi \psi -[ c , \chi] ).
\end{eqnarray}
The fixed point is determined by
$(\phi(1-\phi))=0$.
If this $\phi$ is not matrix valued, then the only nontrivial solution 
which is a smooth function is $\phi=1$.
However, if $N >1$ then a projection 
operator $ P$ which restricts the rank $N$ vector space to dimension $k$
for each point in $M$,
is a solution.
In other words, the solution of $\phi$ is a classification 
map to $G_k (N)$ whose homotopy class classifies the vector bundle.\\

We can construct $S_{pro}$ by 
following the general method of cohomological
gauge theory \cite{Cordes-Moore-Ramgoolam}.
Let us introduce an anti-ghost $\bar{c}$ whose ghost number is $-2$ and 
its BRS partner $\eta$. Then, $S_{pro}$ is given as  
\begin{eqnarray}
S_{pro}=i\int Tr \hat{\delta}  (
( C^{\dagger} \psi ) \bar{c} ) . 
\end{eqnarray}
Here $C^{\dagger}$ is the adjoint operator of $C$. 
$C$ is defined by $\delta_g \phi(x) = i[c , \phi]= C c(x)$  i.e.
$C=i[\makebox{ }, \phi]$.
(Precisely speaking, we define a group action of $U(N)$
for some point $p$ in the principal bundle $P$ over the base manifold $M$.
Then we can define $C$ as the 
differential of the group action on the point $p$;
$C : u(N) \rightarrow T_p P $. 
The image of $C$ is the vertical tangent space of $p$.)
\begin{eqnarray}
S_{pro}=\int Tr \{ 
i[\phi , [c , \phi]]\bar{c}+ 
[\psi , \psi]\bar{c} -[\psi , c ] \eta \} \label{S_pro}
\end{eqnarray}
When we consider the theory near the rank $k$ solution, 
the gauge symmetry $U(N)$ is broken to $U(k)\times U(N-k)$.
Note that 
for a rank $k$ projection operator $\phi$
there are objects $c$ satisfying 
$C^{\dagger}C c= [\phi , [c , \phi]]=\phi c (\phi -1) - (1-\phi)c \phi=0$
i.e. if $c$ is a generator of the gauge group of $U(N-k)\times U(k)$,
then the first term of the right hand side of (\ref{S_pro}) vanishes.
This zero mode causes other types of problems that should be
solved by inserting observables and choosing a good gauge. 
However, to inquire further into this matter would lead us into that 
specialized area, and such a digression would obscure the
outline of our argument. Therefore, we do not go into these details here. 

In the following discussion, $1/C^{\dagger}C$ operates
non-zeromodes and we assume here that there are methods to take care 
of the zero modes.
It is a well known fact of Cohomological gauge theory, that
{}from the $\bar{c}$ equation of motion
$c $ is given as the curvature of the moduli space. 
But this discussion cannot be adopted to our case,
since the non-trivial solution of $\phi$ causes 
symmetry breaking.
The moduli space is the coset space whose equivalent
relation is given by a left gauge symmetry.
\begin{eqnarray}
{\mathcal M}_{k,N}=\{ \phi \ |\  M \rightarrow G_k(N) \}/ 
{\mathcal G}_{k,N},
\end{eqnarray}
where ${\mathcal G}_{k,N}$ is the group of gauge transformations   
with gauge group of $U(N-k)\times U(k)$.
{}From the $\bar{c}$ equation of motion we get
\begin{eqnarray}
c = -\frac{1}{C^{\dagger}C}[\psi , \psi].
\end{eqnarray}
Unlike the usual case, 
we can not regard $c$ as the curvature on the principal bundle whose base
manifold is the moduli space.

Let us consider fermionic zermo-modes of $\chi$ and $\psi$.
Similar to N.C.CohFT and the finite matrix model,
the equations of motion of $\psi$ and $\chi$ without nonlinear terms are 
\begin{eqnarray}
\psi (1-P) -P \psi =0 , \ 
\makebox{and} \ \chi (1-P) -P \chi=0.
\end{eqnarray}
Note that the solution of both equations represents the cotangent 
vector of the solution space of $\phi$.
As far as these equations are concerned, 
the number of zero modes of $\psi$ is equal to 
that of $\chi$ and there is no ghost number anomaly.
After the integration of the nonzeromodes that produces some sign factor
$\epsilon_{k,N}= \pm 1$, the integral of the 
zero-modes is obtained as 
\begin{eqnarray}
{\mathcal E}_{k,N}:=\int_{{\mathcal M}_{k,N}} {\mathcal D}\phi_0 {\mathcal D}\chi_0 
e^{-\int \frac{1}{C^{\dagger} C} [\psi_0, \psi_0][\chi_0, \chi_0]}.
\label{zeromode_gauge}
\end{eqnarray}
Now, recall that our theory has a symmetry that allows 
arbitrary infinitesimal $\phi$ deformation i.e.
$\phi \rightarrow \phi + \delta \phi$, where
$\delta \phi$ is arbitrary infinitesimal 
$N\times N$ Hermite matrix valued scalar field.
This due to the fact that we can regard the BRS exact action as a gauge fixing
action of this local symmetry.
This symmetry means that the partition function is homotopy invariant.
Therefore, the equivalence class of this symmetry corresponds to 
homotopy equivalent class of $\phi$.
So the zero-mode integral (\ref{zeromode_gauge}) is 
summed up by the homotopy class $[M,G_k(N)]$.

Finally, the partition function is given as 
\begin{eqnarray}
Z\sim \sum_{[M,G_k(N)]} \epsilon_{k,N}\ {\mathcal E}_{k,N}
\end{eqnarray}
To interpret this partition function from the point of view of
 classifying homotopy of vector bundles,
we note that
$\phi$ is a classifying map for complex vector bundles 
when $N$ is large enough (see \cite{Bott}).
(Note that there are no non-trivial vector bundles with dimension of 
fiber space larger than $n +2$.)

We introduce the homotopy class $Vect_k(M)= [M , BU(k)]$, where 
\begin{eqnarray}
BU(k)\equiv \bigcup_{m=k+n+1}^{\infty} Gr_k(m)\ ;\  m>k+n, \nonumber
\end{eqnarray}
and consider the case where $N$ is sufficiently large.
Using this, the partition function is represented as 
\begin{eqnarray}
Z\sim \sum_{Vect_k(M)} \epsilon_{k,N}\ {\mathcal E}_{k,N} .
\end{eqnarray}
Note that this homotopy class is related to the
$K'(M)$ group with virtual dimmension 0, where $K'(M)=[M , BU(\infty)]$
(see for example \cite{szabo}).
In particular, when $M$ is connected 
$K(M) ={\mathbb Z} \oplus \tilde{K}$ and $K'=\tilde{K}$.
For stable range $k>\frac{1}{2}\dim M$,
we can put the relation between the homotopy class and 
$K'(M)$ as $K'(M)=[M , BU(k)]$.
Therefore the partition function is proportional to
the sum of $\epsilon_{k,N}\ {\mathcal E}_{k,N}$ over the $K'(M)$
elements for sufficiently large $N$.
This is analogous to the  N.C.CohFT partition function 
which is given by a sum over the elements of the algebraic K-group.
(See also the next subsection.)

To compare with the noncommutative theory with kinetic terms,
we consider the model (\ref{CCoh}) with kinetic terms and
investigate its large scale limit and finite scale case.
The Lagrangian is similar to the N.C.CohFT given in section 4.3; 
\begin{eqnarray}
\mathcal{L}&=&
 {\hat{\delta} }\left( \frac{1}{2} 
 \chi   \left( 2(\phi  (1- \phi )  -\partial_{\mu} B^{\mu}) 
 -iH \right) \right)
 \nonumber \\
 &&
 +\hat{\delta}  \left(\frac{1}{2}
    \chi^{\mu}  ( \partial_{\mu} \phi + B_{\mu} - iH_{\mu})\right) . 
\end{eqnarray}
Since the $U(N)$ gauge symmetry is not our main concern, we break 
gauge symmetry here,
i.e. we do not introduce gauge fields and gauge covariant derivatives.
In the N.C.CohFT case, we took the large $\theta$ limit.
We can perform a similar discussion by scaling
\begin{eqnarray}
g_{\mu \nu} \rightarrow (1+ \epsilon^2) g_{\mu \nu} , \ \ 
g^{\mu \nu} \rightarrow (1- \epsilon^2) g^{\mu \nu}.
\end{eqnarray}
Since the partition function is invariant under this 
transformation, the kinetic terms become irrelevant in the large scale 
limit and $B_{\mu}$ becomes an auxiliary
field. After integration, the theory is equivalent to the one with
the above action (\ref{CCoh}).
This observation is similar to the case of N.C.CohFT in the limit $\theta \rightarrow \infty$.\\

The N.C.CohFT in the previous section is the 
naive extension of
the model dealt with in this section. 
Considering the noncommutative deformation of the model 
in this subsection, we can identify it with the N.C.CohoFT model of 
section 4, after renumbering the U(N) indices and 
Hilbert space indices such that we do not have to distinguish them.
Alternatively, the N.C.CohFT model is obtained by
dimensional reduction to dimension zero and taking the large $N$ limit.

\subsection{$K_0$ and N.C.CohFT}

In this subsection, we disscuss the correspondence with
$K_0$-theory.
As mentioned in section 1, one of our purposes is to construct a 
less sensitive topology than $K$-theory, where the term ``topology" means that  the vacuum expectation value of the field theory which is invariant 
under continuous deformation.
It is natural to expect that our partition function is 
invariant under deformations which do not change the $K$-theory.
In a sense, $\theta$ independence of the partition function
implies this fact.
To see this, we consider the examples of the Moyal plane and the
noncommutative torus.

For the Moyal plane, 
as we saw in the previous section, the partition function (\ref{z2}) 
is expressed as summation over the projection operators 
that are identified by their rank.
The rank of the projection can be identified as
$\tau_0(P_k)=k$ or $\tau_2(P_k)=k $ 
(see for example \cite{Connes} and \cite{Matsuo}). 
Furthermore, the Euler numbers of the Grassmann manifolds 
are determined essentially by $k$ since we take $N \rightarrow \infty$ 
in the end.
Therefore, the partition function is determined by
the $K_0$ data only.\\

Next, we consider the N.C.torus $T^2_\theta$.
The classification by Morita equivalence corresponds to 
the one by $K$-theory and the equivalence is determined by a
noncommutative parameter $\theta$ up to SL(2,${\mathbb Z}$) transformation.
If $T^2_\theta$ and $T^2_{\theta'}$ are Morita equivalent, 
$\theta'$ should be written as
\begin{eqnarray}
\theta'= \frac{a \theta +b}{c \theta + d} \ \  ,\  ad-bc=1 , \  
\ a,b,c,d \in {\mathbb Z}.
\end{eqnarray}
For arbitrary $\theta$ we can transform $T^2_\theta $
to a non-Morita equivalent noncommutative torus by infinitesimal
$\theta $ deformation.
So, the $\theta$ shift changes the $K$-group. 
On the other hand, the model whose action is given by 
(\ref{S_1}) or (\ref{S_1withKinetic})
is invariant under the $\theta$ shift when there is no singular point.
(Note that the models with action (\ref{S_2}) or (\ref{S_2withKinetic})
are not invariant under an arbitrary $\theta$ deformation, but they are 
invariant under 
SL(2,${\mathbb Z}$) transformation.)
At least, if some deformation of noncommutative manifolds does not change the
$K$-theory, we may expect that the partition function of N.C.CohFT
will not change.
This fact implies that the partition function satisfies the
condition we are interested in, that is, 
less sensitive topological invariant than $K$-theory.



\section{ N.C. Cohomological Yang-Mills Theory} 

In this section, Cohomological Yang-Mills theories on noncommutative
manifolds are discussed.
If there is gauge symmetry, the BRS-like symmetry is slightly 
different from (\ref{BRS}), i.e., it is not nilpotent but 
\begin{eqnarray}
\delta^2 =\delta_{g, \theta}, \label{del_g}
\end{eqnarray}
where
$\delta_{g, \theta}$ is gauge transformation operator 
deformed by the star product 
$\mbox{\Large $*$}_{\theta}$.
The partition function of the N.C.CohFT is invariant 
under a change of the noncommutative parameter when the BRS transformation
is nilpotent, since the BRS transformation ${\delta}$ and the
$\theta$ deformation $\delta_{\theta}$ commute.
Conversely, if the definition of the BRS-like operator (\ref{del_g}) 
depends on 
the noncommutative parameter $\theta$, then $\delta$ and 
$\delta_{\theta}$ do not commute;
\begin{eqnarray}
\delta_{\theta} \delta \neq \delta \delta_{\theta} \Rightarrow 
\delta_{\theta} \delta = \delta' \delta_{\theta} ,
\end{eqnarray}
where $\delta'$ is the BRS-like operator that generates
the same transformations as the original BRS-like operator $\delta$ 
without the square
\begin{eqnarray}
{\delta'}^2 =\delta_{g, \theta+\delta \theta}.
\end{eqnarray}
This fact causes some complexity when we want to prove the $\theta$-shift
invariance of N.C.cohomological Yang-Mills theory as mentioned in section 2.\\

Note that the important point of this problem is not the change of the 
nilpotency, but the dependence on the $*$ product in the definition of the 
BRS operator. 
(In fact, we can construct the BRS operator for the cohomological Yang-Mills 
theory as a nilpotent operator \cite{P.R}. )\\

However, we can prove the invariance of the partition function
of cohomological Yang-Mills theory in noncommutative ${\mathbb R}^4$
under the noncommutative parameter deformation 
$$\theta \rightarrow \theta+\delta \theta \ \ , 
(\theta^{\mu \nu}) = 
\left(
\begin{array}{cc|cc}
0& \theta &0&0 \\
-\theta & 0 &0&0 \\
\hline
0&0 &0& \theta  \\
0&0& -\theta &0 
\end{array}
\right) \ .
$$ 
When we consider only the case of noncommutative ${\mathbb R}^4$,
field theories are expressed by the Fock space formalism.
Then the differential operator $\partial_{\mu}$ is expressed by using 
commutation bracket $ -i \theta^{ -1}_{\mu \nu}[ x^{\nu} , \ ] $
and $\int d^D x $ is replaced by $ det (\theta )^{1/2}Tr $.
Therefore the noncommutative parameter deformation is 
equivalent to replacing $ -i \theta^{ -1}_{\mu \nu}[ x^{\nu} , \ ] $
and $ det (\theta )^{1/2}Tr $ by 
$ -i (\theta+\delta \theta)^{ -1}_{\mu \nu}[ x^{\nu} , \ ] $
and $ det (\theta +\delta \theta)^{1/2}Tr $, respectively.
Let us consider 
Donaldson-Witten theory 
(topological twisted ${\mathcal N}=2$ Yang-Mills theory) on
noncommutative ${\mathbb R}^4$ \cite{TFT}.
This theory is constructed by bosonic fields
$(A_{\mu}, H_{\mu \nu} , \bar{\phi},  \phi )$ and 
fermionic fields $(\psi_{\mu} , \eta , \chi_{\mu \nu})$, 
where $(A_{\mu}, H_{\mu \nu} , \bar{\phi} )$
and $(\psi_{\mu} , \eta , \chi_{\mu \nu})$ are supersymmetric
(BRS) pairs.
Using the BRS symmetry, it has been proved that 
partition function and expectation values
have symmetry under the
following reparametrization :
\begin{eqnarray}
A_{\mu} \rightarrow \frac{1}{g} A_{\mu} , \ \ 
\psi_{\mu} \rightarrow \frac{1}{g} \psi_{\mu} & , &
\bar{\phi} \rightarrow \frac{1}{g^2} \bar{\phi} , \ \ 
\eta \rightarrow \frac{1}{g^2} \eta
\nonumber \\
\chi_{\mu \nu} \rightarrow \frac{1}{g^2} \chi_{\mu \nu} & , & 
H_{\mu \nu} \rightarrow \frac{1}{g^2} H_{\mu \nu} , \label{repara}
\end{eqnarray}
where $g \in {\mathbb R}$. 
Taking $g=\theta$, the invariance of Donaldson-Witten
theory under $\theta \rightarrow \theta + \delta \theta$ is proved.

Similarly, we can discuss 
the Vafa-Witten theory
(topological twisted ${\mathcal N}=4$ Yang-Mills theory) on
noncommutative ${\mathbb R}^4$ \cite{Vafa-Witten}.
In this case, there are additional fields 
$(B_{\mu \nu } , c , H_{\mu})$ and $(\psi_{\mu \nu} , \bar{\eta}, \chi_{\mu} )$
, where $(B_{\mu \nu } , c , H_{\mu})$ are bosonic fields, 
$(\psi_{\mu \nu} , \bar{\eta} , \chi_{\mu})$ are fermionic, and they are 
supersymmetric partners.
For these fields, we assign the weight of the above transformation as
\begin{eqnarray}
 B_{\mu \nu } \rightarrow \frac{1}{g} B_{\mu \nu }\ \ 
 , \ \ \psi_{\mu \nu} \rightarrow \frac{1}{g} \psi_{\mu \nu}
 &,&  c \rightarrow \frac{1}{g} c \ \ , \ \ 
\bar{\eta} \rightarrow \frac{1}{g}  \bar{\eta} , \nonumber \\
H_{\mu } \rightarrow \frac{1}{g^2} H_{\mu }\ &,& \ \ 
\chi_{\mu } \rightarrow \frac{1}{g^2} \chi_{\mu }
\end{eqnarray}
with (\ref{repara}), 
then the invariance of Vafa-Witten
theory under $\theta \rightarrow \theta + \delta \theta$ is proved
in the same way as Donaldson-Witten theory.
\\

{}By applying these facts to several physical models,
interesting information can be found.
For example, the partition function of the 
N.C. Cohomological Yang-Mills Theory
on 10-dim Moyal space and the partition function 
of the IKKT matrix model
have a correspondence, since the IKKT matrix model is 
constructed as a dimensional reduction of
super $U(N)$ Yang-Mills theory with large $N$ limit \cite{IKKT}
\cite{IKKT2}.
This dimensional reduction is regarded as a large limit of the 
noncommutative parameter ($\theta \rightarrow \infty$ in section 4).
Taking the large $N$ limit of the matrix model is equivalent to considering 
the Yang-Mills theories on noncommutative Moyal space, i.e.
matrices are regarded as linear transformation of the Hilbert space
caused by the noncommutativity in a similar manner as in the case of the
N.C.CohFT on the Moyal plane.
In particular, the  Noncommutative Cohomological Yang-Mills model
on Moyal space in the large $\theta$ limit 
is almost the same as the model 
of Moore, Nekrasov and Shatashvili \cite{Moore-Nekrasov-Shatashvili}.
Moore et al. showed that the partition function is calculated 
by a cohomological matrix model in \cite{Moore-Nekrasov-Shatashvili}
and related works can be seen in 
\cite{Hirano-Kato,Sugino,Bruzzo-Fucito-Morales-Tanzini}. 
We can be fairly certain that we can reproduce their results
by using N.C.cohomological Yang-Mills theories.

Another example is an application to N=4 d=4 Vafa-Witten theory 
\cite{Vafa-Witten}.
The theory is constructed as a balanced CohFT 
(see \cite{BTFT} and \cite{DPS}).
The partition function of the Vafa-Witten theory
is given by a sum of the Euler numbers of the instanton moduli
space over all instanton numbers, provided the vanishing theorem is true.
Here the vanishing theorem guaranties that the fixed point locus of the 
theory is the instanton moduli space.
On commutative manifolds, one of the conditions for the
vanishing theorem to hold 
is that there is no $U(1)$ instanton.
On the other hand, existence of $U(1)$ instantons is well-known 
in noncommutative Moyal space \cite{Nekrasov1,Nekrasov3},
so it is likely that the $U(1)$ instantons also exist on the other
noncommutative manifolds even if these manifolds do not have 
$U(1)$ instantons before the noncommutative deformation.
Therefore, if we consider the Vafa-Witten theory
on noncommutative manifolds, the $U(1)$ instanton effect
appears as the difference to the commutative manifold case.
The result is a sum of Euler numbers of the instanton moduli spaces
and the moduli spaces are deformed by the $U(1)$ instanton effect.
In this case, we expect that its partition function 
on a commutative manifold is 
computed by a matrix theory calculation like 
\cite{Moore-Nekrasov-Shatashvili}.
{}By comparing this partition function, it is reasonable to 
suppose that the Euler number of the deformed moduli space is given,
and we obtain a partition function 
on a manifold that does not satisfy the vanishing theorem.
Such differences from the CohFT on commutative manifolds 
emphasize that N.C.CohFT is non-trivial, though it is less sensitive
than K-theory.

With these considerations we find many interesting 
subjects to be studied
by using N.C.chomological Yang-Mills theory, and many of them 
will be left for our future work.

\section{Summary}

Let us summarize this article.
We have studied topological aspects of N.C.CohFT and 
matrix models.
At first, we reviewed N.C.CohFT and its properties.
In particular, through this article we have used the property
that the N.C.CohFT has symmetry under an arbitrary 
infinitesimal deformation of the noncommutative parameter.
This symmetry implies that the partition function of N.C.CohFT
is an insensitive ``topological" invariant.
In section 3, 
we introduced a Hermitian finite size matrix model of CohFT and calculated 
its partition function. The calculation was done by using only
topological information of its moduli space. The partition function was given 
as a sum of Euler numbers of Grassman manifolds
with sign, and we showed that the partition function vanishes.
This calculation is the first example
of determining its partition function 
by only moduli space topology of a matrix model.
The scalar field models of N.C.CohFT were discussed in section 4.
The variations of the models are caused by adding kinetic terms or 
topological action that correspond to Connes's Chern character.
The fixed point loci of the scalar fields were given by the set of 
all projection operators on the noncommutative manifold.
{}From the analogy of the finite size matrix model,
we introduced a connection functional in these N.C.CohFT models.
Using the curvature obtained from these connections, 
the partition functions were represented as a sum of Euler numbers of
the set of all projection operators.
This partition function is an example of a new ``topological" invariant,
and the fundamental theorem of Morse theory extended to the 
operator formalism \cite{sako} is connected to the usual local geometry 
by this model.
As a concrete example, we calculated the partition function
of a model including kinetic terms on the Moyal plane.
Through the operator formulation, this calculation boiled down to 
the calculation of the Hermitian finite size matrix model of CohFT
in section 3.
Additionally, to confirm the independence of the noncommutative
parameter of the N.C.CohFT we studied the moduli space for finite
$\theta$.
If the partition function of CohFT is ``topological",
then it should have some relation with K-theory and 
the partition function should not change under a deformation that 
does not change the K-group.
Therefore we investigated the models of CohFT and N.C.CohFT 
from the point of view
of K-theory.
At first, one CohFT was constructed. 
This model and the N.C.CohFT model in section 4 are related 
by dimensional reduction or noncommutative deformation.
The partition function is invariant under scaling and this 
scaling is similar to the $\theta$-shift.
In the large scaling limit, the kinetic terms become irrelevant
and the fixed point loci are given by a classifying map.
The partition function was given by
sum of topological invariants with sign. 
This sum is taken over all the homotopy
equivalent classes of the classifying map of the vector bundle.
The corresponding homotopy class is regarded as $K'$.
{}Comparing the connection between the CohFT model and K-theory
with the relation between 
the N.C.cohomological scalar model and algebraic K-theory, respectively,
we found an analogy.
Furthermore, we studied the correspondence with
the $K_0$-theory for the Moyal plane and the 
noncommutative torus.
It was verified that our partition function is 
invariant under deformations which do not change the $K_0$,
at least for the Moyal plane and 
noncommutative torus. 
These fact implies the existence of a new classification
by the ``topological" invariants.
To clarify the nature of this classification, it is a future work to 
collect examples for which similar calulations  
like the case of the Moyal plane are possible.

Furthermore, we considered the noncommutative cohomological Yang-Mills theory.
The noncommutative parameter independence is non-trivial 
for noncommutative gauge theory, and it is possible to be proved. 
Therefore, we can remove kinetic terms in the large $\theta$ limit
on the Moyal spaces, similar as in the case of the N.C.CohFT 
studied in section 4.
The observations of the N.C.CohFT of scalar models give us 
a general correspondence between N.C.CohFT and matrix models.
An example is the connection between the IKKT matrix model 
and noncommutative cohomological Yang-Mills theory.
{}As another example, we considered the Vafa-Witten theory.
The contribution from noncommutative solitons 
like $U(1)$ instantons
may make the expectation value of N.C.CohFT different 
from the expectation value of CohFT
on a commutative manifold. 
In such case, N.C.CohFT lead to a topological invariant which is different 
from the topological invariant of the commutative case, and which is less sensitive than algebraic K-theory.
In other words, there will be a new nontrivial global 
characterization of the geometry, though its classification is less sensitive
than K-theory. 
It is likely that the Vafa-Witten theory 
is one such example.
A detailed analysis of similar variations for 
noncommutative cohomological Yang-Mills theory corresponding to matrix model
will be carried out in future work.

The unsettled questions are the following.
As we have seen, there is evidence to suggest the
partition function of N.C.CohFT is an insensitive but nontrivial topological invariant.
To clarify this, a more strict topological discussion about N.C.CohFT 
for the general case should be done, since there are many ambiguous problems
concerning the relation to K-theory.
This subject is left as a future issue. \\

\hspace*{-6mm}
{\large\bf  Acknowledgments}\\
We are grateful to Y.Maeda and  H.Moriyoshi for helpful suggestions and
observations.
We also would like to thank Y.Matsuo, M.Furuta, Y.Kametani
for valuable discussions and useful comments.
This work is supported by the 
21st Century COE Program at Keio University (
Integrative Mathematical Sciences:
Progress in Mathematics Motivated by Natural and Social Phenomena ).
The Yukawa Institute for Theoretical Physics at Kyoto University is also 
greatly acknowledged. Discussions during the YITP workshop YITP-W-03-07 on ``Quantum Field Theory 2003'' were useful to complete this work. 
Thanks are due to U.C-Watamura for reading the entire text in its original
 form and collecting the proofs.

\end{document}